\documentstyle[prl,aps,floats,eqsecnum,psfig]{revtex}

\begin{document}
\draft
\def\overlay#1#2{\setbox0=\hbox{#1}\setbox1=\hbox to \wd0{\hss #2\hss}#1
-2\wd0\copy1}
\twocolumn[\hsize\textwidth\columnwidth\hsize\csname@twocolumnfalse\endcsname

\title{Endoscopic Tomography and Quantum-Non-Demolition}

\author{Mauro~Fortunato\cite{mau} and Paolo~Tombesi}
\address{Dipartimento di Matematica e Fisica, Universit\`a di
        Camerino, I-62032 Camerino, Italy \\ and Istituto Nazionale per la
        Fisica della Materia, Unit\`a di Ricerca di Camerino}
\author{Wolfgang~P.~Schleich}
\address{Abteilung f\"ur Quantenphysik, Universit\"at Ulm, D-89069 Ulm,
         Germany}
\date{Received \today}
\maketitle
\begin{abstract}
We propose to measure the quantum state of a single mode of the radiation 
field in a cavity---the signal field---by coupling it via a
quantum-non-demolition Hamiltonian to a meter field in a highly squeezed state.
We show that quantum state tomography on the meter field using balanced
homodyne detection provides full information about the signal state.
We discuss the influence of measurement of the meter on the signal field. 
\end{abstract}
\vskip2pc]
\narrowtext

\section{Introduction}
\label{intro}

How to measure the quantum state of a single mode of the radiation field in a
cavity? Various possibilities~\cite{kn:bard,kn:spec,kn:phwo,kn:frhe,kn:luda,%
kn:kima,kn:leib} offer themselves. However, a straightforward
application of the method of quantum state tomography suggested in
Ref.~\cite{kn:vog} and implemented experimentally in
Refs.~\cite{kn:smith,kn:beck,kn:natu} does not work, since
by coupling the field out of the resonator we change the field state.
In the present paper we propose to couple the field via a
quantum-non-demolition (QND) interaction~\cite{kn:sum} to a meter field on
which we then perform tomography using a balanced homodyne detector.
In this way we combine the idea of probing, that is doing endoscopy on the 
field without taking it out of the cavity, and the tool of tomography and
arrive at the method of endoscopic quantum state tomography.

The goal of the present paper is to obtain information about the full quantum
state of a single mode of the radiation field.
To bring out the physics most clearly
we assume that this field, referred to in the remainder of this article by 
the signal mode, is in a pure quantum state and neglect damping.
We emphasize, however, that the method presented here also applies 
to a signal field described by a density operator. 
In contrast to the method of quantum state
tomography~\cite{kn:vog,kn:smith,kn:beck,kn:natu}
based on homodyne detection, the present technique does not couple
the signal field out of the resonator. In order to measure the signal field
we couple it in a linear way to a meter field. Moreover, we couple both to 
a pump field. This allows us to achieve a quantum-non-demolition Hamiltonian
describing the interaction between the signal and the meter mode. The use 
of a QND-Hamiltonian suggests that one might be able to arrange the scheme 
in such a way as to measure a complete quadrature distribution without 
re-preparing the quantum state. In other words, repeated measurements on the 
meter {\it change} the signal {\it state} but {\it keep} the quadrature 
{\it distribution invariant}. We show that unfortunately this is not the case.
This is closely related to the question if the wave function of a single
quantum system could be measured~\cite{kn:roy}.
Indeed Ref.~\cite{kn:aha} suggests that
the wave function of a single quantum system could be measured by employing
a series of ``protective measurements'' where an {\it a priori} knowledge
of the wave function enables one to measure this wave function and protect it
from changing at the same time. However, Alter and Yamamoto~\cite{kn:orly}
showed that a series of repeated weak quantum non-demolition measurements
gives no information about the wave function of the system. The same
authors~\cite{kn:alter} have also argued that it is not allowed to measure
the full state of a single quantum system. Recently, D'Ariano and
Yuen~\cite{kn:dy} have independently proven the impossibility of measuring
the wave function of a single quantum system. The present intentions are
much less ambitious since, eventually, we do not want to measure the full
state of a single quantum system, but only the quadrature probability
distribution.

The article is organized as follows: In Sec.~\ref{hami} we re-derive the
relevant QND Hamiltonian emphasizing its dependence on the phase of the
pump field which allows us to probe all quadratures of the signal field.
We devote Sec.~\ref{enta} to the calculation of the entangled state of
meter and signal originating from the unitary time evolution due to the
QND Hamiltonian. In Sec.~\ref{cond} we study the influence of the
measurement of the meter on the signal field and in Sec.~\ref{exa} we
consider two special cases: in phase and out of phase measurements.
In Sec.~\ref{meter} we then turn to the question of tomography using a
QND Hamiltonian. In Sec.~\ref{proof} we give a general argument which
shows the impossibility of having a (QND) measurement which simultaneously
keeps the probability distribution unchanged {\em and} gives information
about the measured observable. We conclude in Sec.~\ref{conclu} by
summarizing our main results.
In order to keep the article self-contained we have included all relevant
calculations but have summarized longer ones in Appendices~\ref{dispmeter}
and \ref{quad}. 

\section{QND Hamiltonian}
\label{hami}

In the present section we derive the QND Hamiltonian used in our tomographic 
scheme to couple the signal to the meter field. This treatment brings out 
clearly how the phase of the pump field allows us to probe every quadrature
of the signal.

Our model starts from the Hamiltonian 
\begin{equation}
\hat {H} \equiv i \hbar \chi [\hat {a}_s^\dagger \hat {a}_m^\dagger \hat 
{a}_p - \hat {a}_s \hat {a}_m \hat {a}_p^\dagger] + i\hbar \sigma [\hat {a}_s
^\dagger \hat {a}_m - \hat {a}_s \hat {a}_m^\dagger]\;,
\label{eq:hami}
\end{equation}
where $\hat{a}_s (\hat{a}_s^\dagger)$, $\hat {a}_m (\hat{a}_m^\dagger)$, and
$\hat{a}_p (\hat{a}_p^\dagger)$ denote the annihilation (creation) operators 
of the signal, meter, and pump field, respectively. The parameters $\chi$
and $\sigma$ measure the coupling between the three fields, and the meter and 
signal field, respectively.

\begin{figure}[t]
\centerline{\psfig{figure=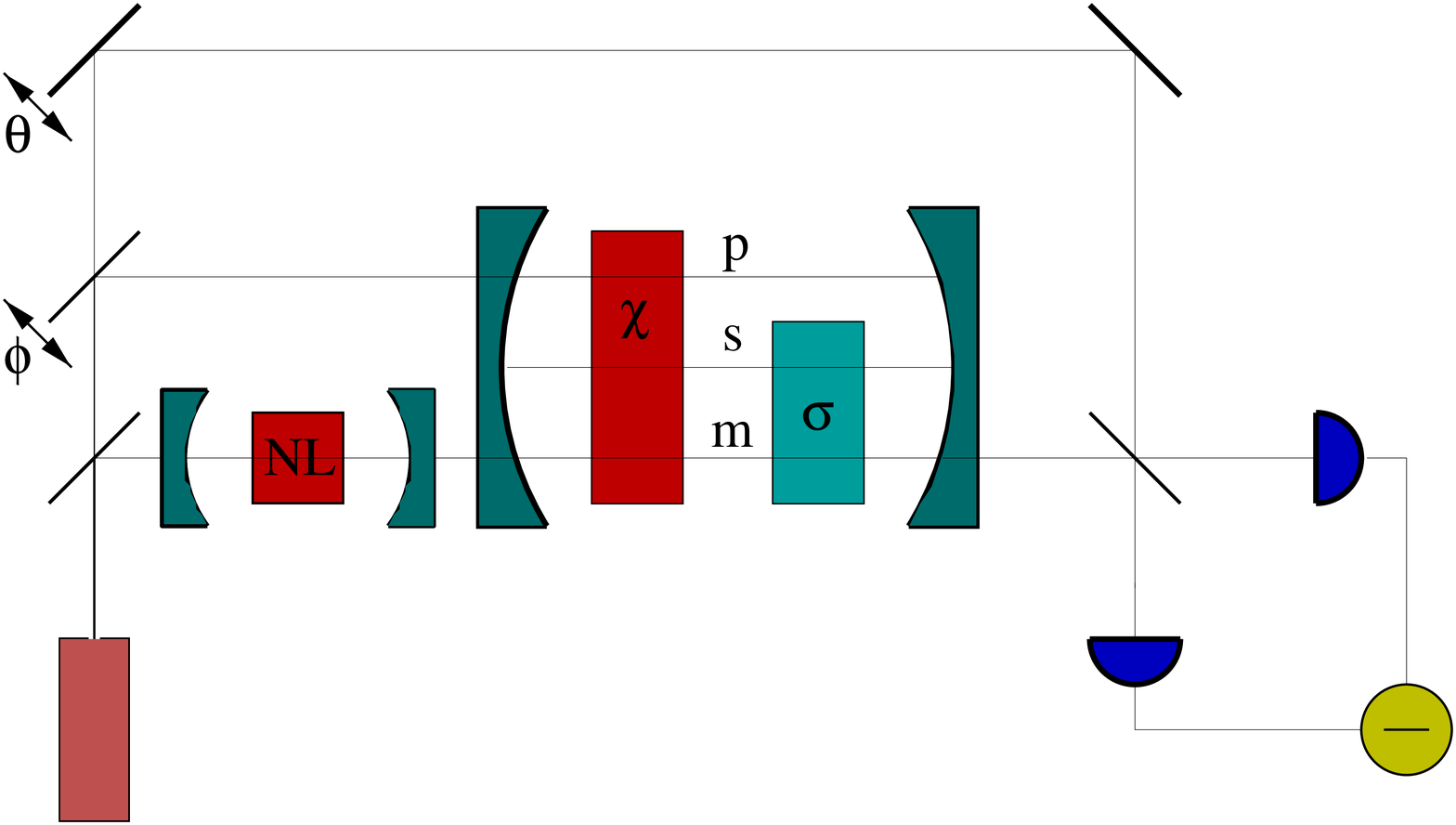,width=3.3in}}
\vspace{0.1cm}
\caption{
Possible scheme for endoscopic tomography. Our goal is to measure
the density operator of the signal $s$ without taking the field out of
the resonator. For this purpose, we couple it via a nonlinear medium of
susceptibility $\chi$ to a pump and a meter field, $p$ and $m$, respectively.
The meter field is in a squeezed state created for example in a separate
resonator by another nonlinear medium NL. The pump field is in a coherent
state of large amplitude $\alpha$ and phase $\phi$. The phase $\epsilon$
of the squeezing parameter is identical to $\phi$.
Apart from the nonlinear coupling between signal and
meter, there is also a linear coupling between the two. When
the susceptibility $\sigma$ is equal to the product $\chi \alpha$, the
effective interaction Hamiltonian for signal and meter is the
product of two quadrature operators of the two fields. In particular, the
two operators are out of phase and their average phase is set by the pump
field.
Using part of the pump field as a local oscillator with phase
$\theta$, we perform a balanced homodyne detection on the meter
coupled out of the resonator. When the homodyne phase $\theta$ is equal
to the pump phase
$\phi$ the interaction displaces the state of the signal field
along the momentum axis and therefore leaves the position distribution
invariant. Moreover, in this case the measured meter distribution is
not influenced by the interaction with the signal. Since we have only
displaced the quantum state of the signal, we have disturbed it in a
controlled way. Unfortunately, we have not obtained any information about it.
When the homodyne phase $\theta$ is
out of phase with the pump phase $\phi$ the interaction changes the
signal as well as the meter field . We therefore obtain information about
the signal field but also disturb it. Consequently, we have to reprepare all
quantum states after each measurement. To reconstruct the quantum state of
the signal using tomography, we record the quadrature distributions of the
meter for all phase angles $0<\theta<\pi$. In these measurements the homodyne
phase $\theta$ has to be locked to the pump phase $\phi$ such that
$\theta=\phi + \pi/2$.
}
\label{fg:scheme}
\end{figure}

A possible scheme of the measurement strategy
suggested in this paper is shown in Fig.~\ref{fg:scheme}. We assume that
the crystal is present in the cavity when we prepare the signal field.
In this case the pump and the meter field are in vacuum states and the
resulting modifications on the signal due to the presence of the crystal can
be easily taken into account.

When the pump field is highly excited we can describe it by a coherent state
of amplitude $\alpha$ and  phase $2\phi$, that is 
\begin{equation}
\hat{a}_p \simeq \alpha e^{2i\phi}\;. 
\label{eq:coh}
\end{equation}
Here we have defined the phase $2\phi$ rather than $\phi$ as to simplify the 
resulting equations. 
It is the variation of this phase $\phi$ of the pump field which allows us to
perform tomography on the signal field. To understand this in more detail we
substitute the coherent state approximation, Eq.~(\ref{eq:coh}), of the pump
field into the Hamiltonian, Eq.~(\ref{eq:hami}), and find after minor algebra
\begin{equation}
\hat{H} = 2\hbar \sigma \hat x_s (\phi + \pi/2) \cdot \hat{x}_m (\phi)\;.
\label{eq:cohami}
\end{equation}
Here we have arranged the strength $\alpha$ of the pump field such that
$\chi\alpha = \sigma$. Moreover, we have introduced the quadrature operators
\begin{equation}
\hat{x}_j (\theta) \equiv \frac{1}{\sqrt{2}} \left(\hat{a}_j e^{-i\theta} +
\hat{a}_j^\dagger e ^{i\theta}\right)
\end{equation}
of the signal $(j=s)$ and the meter $(j=m)$ mode at the angle $\theta$.

Note that due to the special choice $\chi\alpha = \sigma$ of the pump field
we have achieved an interaction between the signal and the meter which couples
the quadrature operator $\hat{x}_m (\phi)$ of the meter at phase angle 
$\phi$ to the out-of-phase quadrature operator $\hat{x}_s(\phi + \pi/2)$
of the signal. Such Hamiltonians have been studied
extensively~\cite{kn:als,kn:kim,kn:per,kn:yur,kn:schl,kn:hil,kn:tom}
in the context of quantum non-demolition measurements. In the present paper we
analyze how such a Hamiltonian can be used to measure the quantum state
of the signal field. We note that according to the QND Hamiltonian
Eq.~(\ref{eq:cohami}) a measurement of the meter at a fixed phase $\phi$
of the pump field provides information about the signal in the out of
phase quadrature. By varying the phase $\phi$ of the pump field we can
probe in this way all quadratures of the signal. We conclude this section
by noting that we can achieve a measurement of the meter quadrature
operator by a homodyne measurement of the meter mode.

\section{Entanglement}
\label{enta}

We now calculate the combined state $|\Psi\rangle$ of signal and meter 
obtained from the QND interaction Hamiltonian, Eq.~(\ref{eq:cohami}).

When we couple the signal and meter mode prepared initially in the states 
$|\psi_s\rangle$ and $|\psi_m\rangle$ we find the quantum state
\begin{eqnarray}
|\Psi(t) \rangle &=& \exp (-i\hat{H}t/\hbar)|\psi_m\rangle|\psi_s\rangle
\nonumber \\
 & = & \exp[-2i\sigma t\hat{x}_s(\phi + \pi/2)
\hat{x}_m(\phi)]|\psi_m\rangle|\psi_s\rangle
\label{eq:evol}
\end{eqnarray}  
for the combined system after the interaction time $t$. This time is
determined by the decay time of the cavity. 

To evaluate the above expression we expand the initial signal state in
quadrature states $|x_s(\phi + \pi/2)\rangle$ of the phase angle
$\theta_s\equiv \phi + \pi/2$, that is
\begin{equation}
|\psi_s\rangle = \int\limits_{-\infty}^\infty dx_s\, \psi_s(x_s; \phi + \pi/2)
|x_s (\phi + \pi/2)\rangle\;.
\label{eq:exp}
\end{equation}
We emphasize that this representation and, in particular, the wave function
$\psi_s(x_s; \phi + \pi/2) \equiv \langle x_s(\phi + \pi/2)|\psi_s\rangle$
depend crucially on the angle $\theta_s$.

We substitute the expression Eq.~(\ref{eq:exp}) for the signal state into
Eq.~(\ref{eq:evol}), use the eigenvalue equation
\begin{equation}
\hat x_s (\theta)|x_s(\theta)\rangle = x_s | x_s(\theta)\rangle
\label{eq:eig}
\end{equation}
for the signal quadrature state $|x_s(\theta)\rangle$ at angle $\theta$, and
arrive at the combined state
\begin{eqnarray}
|\Psi(t) \rangle = \int\limits_{-\infty}^\infty & dx_s & \psi_s (x_s; \phi +
\pi/2)|x_s(\phi + \pi/2)\rangle
\nonumber \\
 & \times & \exp[-2i\sigma t x_s \hat{x}_m (\phi)]|\psi_m\rangle
\label{eq:comb}
\end{eqnarray}
of signal and meter.

To find the action of the exponential operator in Eq.~(\ref{eq:comb}) on
the meter state $|\psi_m\rangle$ it is convenient to expand $|\psi_m\rangle$
in quadrature states $|x_m(\theta)\rangle$ of the meter at the angle
$\theta$, that is
\begin{equation}
|\psi_m\rangle = \int\limits_{-\infty}^\infty dx_m\, \psi_m (x_m; \theta)
|x_m (\theta)\rangle\;,
\label{eq:expan}
\end{equation}  
where $\psi_m (x_m;\theta)\equiv \langle x_m(\theta)|\psi_m\rangle$
denotes the wave function of the meter state at the angle $\theta$.
Note that this angle is still arbitrary and is not necessarily
identical to the angle $\phi$ in the Hamiltonian. According to the
Appendices~\ref{dispmeter} and \ref{quad} we find
\begin{eqnarray}
 & \exp &[-i(2\sigma tx_s)\hat x_m(\phi)]|\psi_m \rangle
\nonumber \\
 & = & \int\limits_{-\infty}^\infty dx_m\exp[-i\gamma(x_s,x_m;\theta-\phi)]
\label{eq:action} \\
 & \times & \psi_m[x_m - 2\sigma tx_s\sin (\theta-\phi);\theta]
 |x_m(\theta)\rangle\;,
\nonumber
\end{eqnarray}
where
\begin{eqnarray}
\gamma(x_s, x_m; \theta-\phi) & \equiv & (\sigma t x_s)^2\sin[2(\theta-\phi)]
\nonumber \\
 & & + 2\sigma t x_s x_m \cos(\theta-\phi)
\label{eq:gamma}
\end{eqnarray}
denotes the phase accumulated due to the interaction. 

Hence the combined quantum state reads
\begin{eqnarray}
|\Psi(t)\rangle &=& \int\limits_{-\infty}^\infty dx_s
\int\limits_{-\infty}^\infty dx_m \psi_s(x_s; \phi + \pi/2)
\nonumber\\
& & \quad \quad \times \psi_m [x_m - 2\sigma tx_s \sin (\theta - \phi); \theta]
\nonumber\\
& & \quad \quad \times \exp[-i\gamma(x_s,x_m; \theta-\phi)]
\nonumber\\
& & \qquad \times |x_s (\phi + \pi/2)\rangle| x_m (\theta) \rangle\;.
\label{eq:combstate}
\end{eqnarray}
We note that due to the coupling between the meter and the signal via the 
Hamiltonian Eq.~(\ref{eq:cohami}), the meter wave function
$\psi_m (x_m;\theta)$ at the angle $\theta$ gets shifted by an amount
$\delta x_m \equiv 2\sigma tx_s \sin (\theta - \phi)$.
This shift is proportional to the interaction strength $\sigma t$, the
signal variable $x_s$ and the sine of the angle $\theta - \phi$.

\section{Signal state conditioned on meter measurement}
\label{cond}

In the preceding section we have calculated the entangled state
$|\Psi \rangle$, Eq.~(\ref{eq:combstate}), of the combined system.
In the present section we show how a measurement of the meter influences
the state of the signal.
In particular, we use the Wigner function approach to discuss the properties
of the signal state conditioned on a quadrature measurement of the meter
variable. Here we first consider an arbitrary quadrature state of phase angle
$\theta$ and then in Sec.~\ref{exa} focus the discussion on two special cases.

According to Eq.~(\ref{eq:combstate}) the conditioned state
\begin{equation}
|\psi_s^{(c)}\rangle=\frac{1}{\sqrt{W(x_m)}}\langle x_m(\theta)|\Psi(t)\rangle
\label{eq:condstate}
\end{equation}
of the signal given that our quadrature measurement at angle $\theta$ has
provided the value $x_m$ reads
\begin{equation}
|\psi_s^{(c)} \rangle = \int\limits_{-\infty}^\infty\, dx_s \psi_s (x_s; \phi
+ \pi/2) f (x_s|x_m)| x_s (\phi + \pi/2)\rangle\;,
\label{eq:estra}
\end{equation}
where the filter function 
\begin{eqnarray}
f (x_s| x_m) &=& \frac{1}{\sqrt{W(x_m)}} \psi_m [x_m - 2\sigma tx_s
\sin (\theta- \phi); \theta]
\nonumber\\
& \times & \exp[-i \gamma (x_s, x_m; \theta - \phi)]
\label{eq:filt}
\end{eqnarray}
originates from the interaction of the signal with the meter. The probability 
$W(x_m)$ of finding the meter variable $x_m$ follows from the normalization
condition 
\begin{equation}
1 = \langle \psi_s^{(c)}| \psi_s^ {(c)} \rangle\;,
\label{eq:norm}
\end{equation}
that is
\begin{eqnarray}
W (x_m) = \int\limits_{-\infty}^\infty & dx_s &|\psi_s (x_s; \phi + \pi/2)|^2
\nonumber\\
 & \times &|\psi_m[x_m - 2\sigma tx_s \sin (\theta - \phi); \theta]|^2\;.
\label{eq:dostar}
\end{eqnarray}
Equation (\ref{eq:estra}) clearly shows how the measurement of the meter
influences the quantum state of the signal: The filter function determined
by the wave function of the meter selects those parts of the signal wave
function that are entangled with the corresponding parts in the meter.
To study this in more detail we now calculate the Wigner function~\cite{kn:wig}
\begin{eqnarray}
W_s^{(c)} (x_s, p_s| x_m) &=& \frac{1}{2\pi} \int\limits_{-\infty}^\infty dy 
\, e^{ip_sy}\langle x_s - y/2| \psi_s^{(c)}\rangle
\nonumber\\ 
& & \qquad \quad \times \langle \psi_s^{(c)}|x_s + y/2 \rangle
\label{eq:wigsigcon}
\end{eqnarray}
of the signal state conditioned on the measured meter value $x_m$. For the sake
of simplicity we have suppressed the angle $\phi + \pi/2$ at the quadrature
states $|x+y/2\rangle$ and $|x-y/2\rangle$. Substituting the state
$|\psi_s^{(c)}\rangle$, Eq.~(\ref{eq:estra}), into this expression we arrive at
\begin{eqnarray}
&W&_s^{(c)}(x_s,p_s| x_m)=\frac{1}{2\pi} \int\limits_{-\infty}^\infty
dy \, e^{ip_sy} \psi_s (x_s -y/2)
\nonumber \\
& \times &  \psi_s^\ast (x_s + y/2) f(x_s -y/2| x_m) f^\ast(x_s + y/2| x_m)\;.
\label{eq:wigpar}
\end{eqnarray}
We express the integral as the convolution~\cite{kn:rel}
\begin{eqnarray}
& & W_s^{(c)} (x_s, p_s| x_m) =
\nonumber \\
& & \qquad \int\limits_{-\infty}^\infty dp'\, W_s (x_s, p_s - p')
W_f (x_s, p'| x_m)
\label{eq:conv}
\end{eqnarray}
between the Wigner function  
\begin{eqnarray}
& & W_s (x_s, p_s) = 
\nonumber \\
& & \qquad \frac{1}{2\pi} \int\limits_{-\infty}^\infty dy \, e^{ip_sy}
\psi_s (x_s -y/2) \psi_s^\ast (x_s + y/2)
\label{eq:wig1}
\end{eqnarray}
of the original signal state and the Wigner function
\begin{eqnarray}
& & W_f (x_s, p_s| x_m) =
\nonumber \\
& &  \frac{1}{2\pi} \int\limits_{-\infty}^\infty dy \, e^{ip_sy}
f(x_s - y/2| x_m) f^\ast (x_s + y/2| x_m)
\label{eq:wig2}
\end{eqnarray}
of the filter provided by the meter measurement. We can easily prove
Eq.~(\ref{eq:conv}) by substituting the expressions
Eqs.~(\ref{eq:wig1}) and (\ref{eq:wig2}) into Eq.~(\ref{eq:conv}),
interchanging the integrations and performing one of them using the
resulting delta function. We then indeed recover the integral
(\ref{eq:wigpar}).

If we substitute the filter function Eq.~(\ref{eq:filt}) into the Wigner
function Eq.~(\ref{eq:wig2}), after minor algebra we obtain 
\begin{eqnarray}
& & W_f (x_s, p_s| x_m) = \frac{1}{W(x_m)} \frac{1}{2\pi}
\int\limits_{-\infty}^\infty dy
\nonumber \\
& &\times \exp\{iy [p_s + 2(\sigma t)^2 x_s \sin [2(\theta-\phi)]
+ 2\sigma tx_m\cos(\theta-\phi)]\}
\nonumber \\
& & \times \psi_m [x_m - 2\sigma t\sin (\theta - \phi)(x_s - y/2)]
\nonumber \\
& &\times \psi_m^\ast [x_m - 2\sigma t\sin (\theta - \phi)(x_s + y/2)]\;.   
\label{eq:wigcomp}
\end{eqnarray}
When we introduce in the last integral the new integration variable
$\bar{y} \equiv 2\sigma t \sin (\theta -\phi) y$, and in the convolution
Eq.~(\ref{eq:conv}) the integration variable
$p \equiv p'/ [2 \sigma t \sin (\theta- \phi)]$,
the Wigner function of conditional state
\begin{eqnarray}
 W_s^{(c)} (x_s, p_s|x_m) & = & \int\limits_{-\infty}^\infty dp W_s 
[x_s, p_s - 2\sigma t\sin(\theta-\phi)p]
\nonumber \\
& & \qquad \times W_f (x_s, p|x_m)
\label{eq:wigfuncon}
\end{eqnarray}
is the convolution of the Wigner function $W_s$ [Eq.~(\ref{eq:wig1})] of the
signal state and the Wigner function
\begin{eqnarray}
W_f (x_s,&p&|x_m) = \frac{1}{W(x_m)} \frac{1}{2\pi} 
\int\limits_{-\infty}^\infty d\bar{y}
\nonumber \\
&\times& \exp[i\bar{y} (p+2\sigma t x_s \cos (\theta-\phi) + x_m \cot 
(\theta-\phi))]
\nonumber \\
&\times& \psi_m(x_m - 2 \sigma t x_s \sin (\theta-\phi) + \bar{y}/2)
\nonumber \\
&\times& \psi_m^\ast(x_m - 2\sigma t x_s \sin (\theta - \phi) - \bar{y}/2)
\label{eq:wigconv}
\end{eqnarray} 
of the filter function. We express the latter in terms of the Wigner function
\begin{eqnarray}
& & W_m (x_m, p_m) =
\nonumber \\
& & \frac{1}{2 \pi} \int\limits_{-\infty}^\infty dy \, e^{ip_my}
\psi_m (x_m -y/2) \psi_m^\ast (x_m +y/2)
\label{eq:wigmet}
\end{eqnarray}
of the meter via the relation
\begin{eqnarray}
W_f(x_s&,&p|x_m)=\frac{1}{W(x_m)}W_m [x_m - 2\sigma tx_s\sin(\theta-\phi); 
\nonumber \\
p&&+2\sigma t x_s \cos(\theta -\phi)+x_m \cot(\theta - \phi)]\;.
\label{eq:rel}
\end{eqnarray}

\section{Special examples for conditioned signal states}
\label{exa}

Whereas in the discussion of Sec.~\ref{cond} the angle $\theta$ of
the meter quadrature is still arbritrary, we concentrate in the present
section on two distinct cases: We choose (i) $\theta = \phi$, that is
we measure in phase and (ii) $\theta = \phi + \pi/2$, that is out of phase
measurement.

\subsection{In phase measurement}
\label{inphase}

If we choose the angle $\theta$ of the meter quadrature to be identical to 
$\phi$, the state $|\Psi \rangle$, Eq.~(\ref{eq:combstate}), of the complete
system reduces to
\begin{eqnarray}
|\Psi(t)\rangle &=& \int\limits_{-\infty}^\infty\, dx_s
\int\limits_{-\infty}^\infty dx_m\psi_s(x_s;\phi+\pi/2)\psi_m(x_m;\phi)
\nonumber \\
& & \times \exp(-i2\sigma tx_s x_m)|x_s(\phi + \pi/2)\rangle|x_m(\phi)
\rangle\;.
\label{eq:redstate}
\end{eqnarray}
Here we have made use of the phase $\gamma=2\sigma tx_sx_m$,
Eq.~(\ref{eq:gamma}), for $\theta = \phi$. Note that this expression
also follows immediately from the Hamiltonian Eq.~(\ref{eq:cohami}) and
the expansions Eqs.~(\ref{eq:exp}) and (\ref{eq:expan}) of the meter and
signal states.
We emphasize that in this case the meter wave function is not shifted.
Nevertheless, the two states are still entangled via the exponential,
Eq.~(\ref{eq:action}).
Since the shift $\delta x_m$ vanishes, the probability  
\begin{equation}
W(x_m) = |\psi_m|^2 \int\limits_{-\infty}^\infty
         dx_s|\psi_s(x_s)|^2= |\psi_m|^2 \;,
\label{eq:prob}
\end{equation}
of finding the meter variable $x_m$ following from Eq.~(\ref{eq:dostar})
for $\theta = \phi$ is identical to the initial probability of the meter,
that is
\begin{equation}
W(x_m)=|\psi_m(x_m)|^2\;.
\label{eq:inprob}
\end{equation}
Here we have used the fact that the original signal wave function is
normalized. Hence, up to an overall phase $\mu_m$ determined by the meter
wave function $\psi(x_m)=|\psi(x_m)|\exp[i\mu(x_m)]$, we find from 
Eq.~(\ref{eq:filt}) the filter function $f(x_s|x_m)=\exp(-i2\sigma t x_s x_m)$,
and from Eq.~(\ref{eq:estra}) the conditioned signal state
\begin{equation}
|\psi_s^{(c)}\rangle = \int\limits_{-\infty}^{\infty}\, dx_s\psi_s(x_s)\exp
(-i2\sigma tx_s x_m)|x_s (\phi + \pi/2)\rangle\;.
\label{eq:stern}
\end{equation}
Note that the measurement of the meter has indeed changed the {\it state} of
the system but did not alter the probability 
\begin{eqnarray}
W(x_s) &=& |\langle x_s|\psi_s^{(c)}\rangle|^2
        = |\psi_s(x_s)\exp(-i2\sigma t x_s x_m)|^2
\nonumber\\
       &=& |\psi_s(x_s)|^2
\label{eq:probab}
\end{eqnarray}
of finding the signal variable $x_s$. This effect of the meter measurement
comes out most clearly in the Wigner function $W_s^{(c)}$ of the conditioned
system state, Eq.~(\ref{eq:conv}). From Eq.~(\ref{eq:wigcomp}) we realize
that for $\theta = \phi$ the Wigner function of the filter reduces to a delta
function in the momentum shift, that is
\begin{equation}
W_f(x_s, p_s| x_m) = \delta (p_s+2\sigma tx_m)\;,
\label{eq:filwig}
\end{equation}
and the Wigner function following from the convolution Eq.~(\ref{eq:conv})
reads
\begin{equation}
W_s^{(c)} (x_s,p_s|x_m) = W_s(x_s,p_s+2\sigma tx_m)\;.
\label{eq:wigfol}
\end{equation}
Hence, the measurement has left untouched the shape of the original state
represented here by the Wigner function but has moved it along the momentum
axis by an amount of $2\sigma tx_m$. Consequently, the measurement did not
change the probability distribution in the conjugate variable, namely the
$x_s$ variable. We note, however, that in this way we cannot gain information
about the signal since according to Eqs.~(\ref{eq:prob}) and (\ref{eq:inprob})
the probability distribution $W(x_m)$ of measuring the variable $x_m$ is
identical to the original distribution.

This finding is actually a rather general result. In fact, it can be
rigorously shown~\cite{kn:oalt} that a (QND) measurement which does not
change the probability density of the observable which is being measured
on a single quantum system gives no information about the measured observable.
Its proof, restricted for clarity to the model considered here, can be found
in Sec.~\ref{proof}.

\subsection{Out of phase measurement}
\label{outofphase}

We now turn to the case of $\theta = \phi + \pi/2$. In this case the shift
$\delta x_m = 2\sigma tx_s$ in the meter wave function is maximal and
according to Eq.~(\ref{eq:gamma}) the phase $\gamma$ vanishes. Hence, the
combined state
\begin{eqnarray}
&\null&|\Psi(t)\rangle = \int\limits_{-\infty}^\infty dx_s
\int\limits_{-\infty}^\infty dx_m \psi_s(x_s; \phi + \pi/2)
\label{eq:costate} \\
&\times&\!\psi_m (x_m-2\sigma tx_s;\phi + \pi/2)\,
          |x_s(\phi + \pi/2)\rangle\,|x_m(\phi + \pi/2)\rangle
\nonumber
\end{eqnarray}
is an entangled state in which the entanglement between the meter and signal
is due to the shift of the meter. In contrast to the discussion of
Sec.~\ref{inphase} we can now deduce properties of the signal from the shift
of the meter wave function. Unfortunately, we cannot simultaneously keep
the probability distribution $W(x_s)=|\psi_s(x_s)|^2$ of the original
signal state invariant, in accordance with the discussion at the end of
Sec.~\ref{inphase} (see Sec.~\ref{proof}).
Indeed, we find from Eqs.~(\ref{eq:estra}) or (\ref{eq:stern}) the conditional
state 
\begin{equation}
|\tilde{\psi}_s ^{(c)} \rangle = \frac{1}{\sqrt{\tilde{W}(x_m)}} \; \int
\limits_{-\infty}^\infty \!\!dx_s \, \psi_s (x_s) \psi_m (x_m - 2 \sigma t x_s)
\,|x_s \rangle
\label{eq:contilde}
\end{equation}
of the system given the meter measurement at phase $\phi+\pi/2$ has 
provided the value $x_m$. The probability
\begin{equation}
\tilde{W}(x_m) = \int\limits_{-\infty}^\infty dx_s |\psi_s (x_s) |^2|
\psi_m (x_m - 2 \sigma tx_s)|^2
\label{eq:wtilde}
\end{equation}
of finding the meter value $x_m$ following from Eq.~(\ref{eq:dostar}) is now a 
convolution of the system and the meter function. In Sec.~\ref{tomo} we will
use this relation to perform tomography on the system. However, in the present
section we focus on how the measurement influences the signal state.
We note that in contrast to the discussion of Sec.~\ref{inphase} the meter
measurement has changed the conditional distribution
\begin{eqnarray}
\tilde{W}_s^{(c)}&(&x_s| x_m)=|\langle x_s| \psi_s ^{(c)} \rangle| ^2
\nonumber \\
=&& \! |\psi_s (x_s)|^2 \frac{| \psi_m (x_m - 2 \sigma t x_s)|^2}{\int d x_s
| \psi_s (x_s)|^2 |\psi_m (x_m - 2 \sigma t x_s)|^2}\;.
\label{eq:changed}
\end{eqnarray}     
of finding the signal variable $x_s$ given a measurement of the meter has 
provided $x_m$. Moreover, the Wigner function of the conditional system state
is now given by
\begin{eqnarray}
P^{(W)} (x_s,p_s|x_m) &=& \frac{1}{\tilde{W} (x_m)}
\int\limits_{-\infty}^{\infty} dy\,
e^{ip_sy}\psi_s^\ast\left(x_s+\frac{y}{2}\right)
\nonumber\\
&\times& \psi_s\left(x_s-\frac{y}{2}\right)
                 \psi_m (x_m - 2 \sigma tx_s + \sigma ty)
\nonumber\\
&\times& \psi_m^\ast (x_m - 2 \sigma tx_s - \sigma ty)\;.
\label{eq:chwigfun}
\end{eqnarray}
This Wigner function can again be expressed as the convolution
\begin{equation}
P^{(W)} (x_s,p_s|x_m)=\int\limits_{-\infty}^{\infty} dp'\, W_s(x_s,p_s-p')
W_f(x_s,p'|x_m)\;,
\label{eq:agacon}
\end{equation}
where $W_s(x_s,p_s)$ is given by Eq.~(\ref{eq:wig1}), and $W_f(x_s,p_s|x_m)$
this time reads
\begin{eqnarray}
W_f(x_s,p_s|x_m)&=&\frac{1}{2\pi \tilde{W}(x_m)}\int\limits_{-\infty}^{\infty}
dy\, e^{ip_sy}
\nonumber \\
& & \times \psi_m\left[x_m-2\sigma t\left(x_s-\frac{y}{2}\right)\right]
\nonumber \\
& & \times \psi_m^\ast\left[x_m-2\sigma t\left(x_s+\frac{y}{2}\right)\right]\;.
\label{eq:wigfthis}
\end{eqnarray}

If we now change the variables $\bar{y}=2\sigma ty$ in Eq.~(\ref{eq:wigfthis})
and $p=p'/2\sigma t$ in Eq.~(\ref{eq:agacon}), we can rewrite
Eq.~(\ref{eq:agacon}) as the convolution
\begin{equation}
P^{(W)}(x_s,p_s|x_m)=\int\limits_{-\infty}^{\infty} dp
W_s(x_s,p_s-2\sigma tp) W_f(x_s,p|x_m)
\label{eq:agaconaga}
\end{equation}
between the Wigner function of the signal state and the filter Wigner function
\begin{eqnarray}
W_f(x_s,p|x_m)&=&\frac{1}{2\pi \tilde{W}(x_m)}\int\limits_{-\infty}^{\infty}
dy\, e^{ipy}
\nonumber \\
& & \times \psi_m\left[x_m-2\sigma tx_s+\frac{y}{2}\right]
\nonumber \\
& & \times \psi_m^\ast\left[x_m-2\sigma tx_s-\frac{y}{2}\right]\;.
\label{eq:filterwig}
\end{eqnarray}
The latter can be expressed in terms of the Wigner function of the meter
[Eq.(\ref{eq:wigmet})] via the relation
\begin{equation}
W_f(x_s,p|x_m)=\frac{1}{\tilde{W}(x_m)}W_m(x_m-2\sigma t x_s;p)\;.
\label{eq:viatherel}
\end{equation}
Now, in contrast to Sec.~\ref{inphase}, the filter Wigner
function~(\ref{eq:viatherel}) does not reduce to a delta function, and
therefore the Wigner function of the conditional signal state is not
identical to the original one any more. This is indeed the effect of the
measurement. This time, however, as we shall see in Sec.~\ref{tomo}, we
can gain information about the signal.

\section{Meter wave function}
\label{meter}

We continue considering the meter measurement at an angle
$\theta = \phi + \pi/2$ but discuss two extreme cases:
(i) The meter wave function is broad  compared to the signal wave function and
(ii) the meter wave function is extremely narrow.
In the first case we do not change the signal state appreciably but can only
learn about the lowest moments of the signal distribution. In contrast, the
second way of making a measurement destroys the state but repeated
measurements on an ensemble of systems all prepared in an identical way
allow us to reconstruct the signal state using tomographic cuts.

\subsection{Weak measurements}
\label{weak}

Since $\psi_m$ is broad compared to $\psi_s$ we can evaluate $\psi_m$ at 
some characteristic value of $x_s$, such as $\langle x_s \rangle$. In 
this case the conditional state, Eq.~(\ref{eq:contilde}), reduces to 
\begin{equation}
|\psi_s^{(c)} \rangle \equiv \int\limits_{-\infty}^{\infty}
dx_s\, \psi_s (x_s)| x_s \rangle\;,
\label{eq:weakcon}
\end{equation}
and the probability 
\begin{equation}
\tilde{W}(x_m) \equiv |\psi_m (x_m+2\sigma t\langle x_s \rangle)|^2
\label{eq:weakprob}
\end{equation}
is the original meter probability shifted by an amount
$2\sigma t\langle x_s\rangle$. Hence, when this shift
$2\sigma t\langle x_s \rangle$ is larger than the width of
$W_m (x_m) = |\psi_m (x_m)|^2$, we can 
learn about $\langle x_s \rangle$. As seen from Eq.~(\ref{eq:weakcon}),
in this case the state of the signal mode does not change appreciably.

\subsection{Tomographic measurements}
\label{tomo}

Optical homodyne tomography~\cite{kn:vog,kn:smith,kn:beck,kn:natu,kn:kien}
is a method for obtaining
the Wigner function (or, more generally~\cite{kn:dmp,kn:gmd,kn:sch}, the matrix
elements of the density operator in some representation) of the electromagnetic
field, preparing the field again in the same state after each measurement.
It therefore consists of an ensemble of repeated measurements of one quadrature
operator for different phases relative to the local oscillator of the homodyne
detector. However, the method first employed in Ref.~\cite{kn:smith} needs a
smoothing procedure, because, in order to reconstruct the Wigner function one
has to perform an integral involving the marginal probability distribution
of homodyne measurement~\cite{kn:vog}. This was indeed performed in
Refs.~\cite{kn:smith,kn:beck} by methods which are standard in tomographic
imaging~\cite{kn:nat}.

In the present section we show that it is possible to perform tomography
on the meter mode to obtain information about the signal state. To this
end, we recall Eq.~(\ref{eq:wtilde})
\begin{equation}
\tilde{W}(x_m) = \int\limits_{-\infty}^\infty dx_s |\psi_s (x_s) |^2|
\psi_m (x_m - 2 \sigma tx_s)|^2\;,
\label{eq:tildew}
\end{equation}
which gives the marginal distribution of the meter (probability distribution
of the results of the measurements of $\hat{x}_m$) in the case of out of
phase measurements. Let us assume that the meter wave function is extremely
narrow, that is the meter is initially in a highly squeezed state, for example
a squeezed vacuum $|0,\xi\rangle$, where $\xi=r e^{i\epsilon}$ is the
squeezing parameter. Then, according to
Eq.~(\ref{eq:tildew}), the marginal distribution $\tilde{W}(x_m)$ is given by
a convolution of the modulus square of the signal wave function with a narrow
Gaussian
\begin{eqnarray}
|\psi_m(x_m&-&2\sigma tx_s)|^2=\frac{1}{\protect\sqrt{\pi}\cosh
r(1-e^{i\epsilon} \tanh r)}
\label{eq:psivac} \\
&\times&\exp\left\{-\left[\frac{1+e^{i\epsilon}\tanh r}{1-
e^{i\epsilon}\tanh r}\right](x_m-2\sigma tx_s)^2\right\}\;.
\nonumber
\end{eqnarray}

Now, if the modulus $r$ of the squeezing parameter is large enough, the
Gaussian~(\ref{eq:psivac}) approaches a delta function in the meter and
signal variables
\begin{equation}
|\psi_m(x_m-2\sigma tx_s)|^2\longrightarrow \frac{1}{|2\sigma t|}
\delta\left(x_s-\frac{x_m}{2\sigma t}\right)\;,
\label{eq:appdel}
\end{equation}
and Eq.~(\ref{eq:tildew}) reduces to
\begin{mathletters}
\label{eq:double}
\begin{eqnarray}
\tilde{W}(x_m)&\cong&\frac{1}{|2\sigma t|}\int\limits_{-\infty}^{\infty}
dx_s\, |\psi_s(x_s)|^2\delta\left(x_s-\frac{x_m}{2\sigma t}\right)
\label{eq:doublea} \\
&=& \frac{1}{2\sigma t} \left|\psi_s\left(\frac{x_m}{2\sigma t}\right)\right|^2
=\frac{1}{2\sigma t} W\left(\frac{x_m}{2\sigma t}\right)\;.
\label{eq:doubleb}
\end{eqnarray}
\end{mathletters}
Hence, by measuring the probability distribution $\tilde{W}(x_m)$ of the
outcomes of the meter variable $x_m$ (for example via balanced homodyne
detection performed on the meter field) we indirectly obtain the probability
distribution $W(x_s)$, up to a rescaling given by the factor $2\sigma t$.
However, from Eq.~(\ref{eq:contilde})
it is clear that in this case the signal wave function is changed, and
therefore we need to prepare the signal field again in the same state after
each measurement. This is what is usually done in quantum optical
tomography~\cite{kn:smith,kn:beck,kn:natu}.

The advantage of the present scheme is that we perform an indirect measurement:
We do not detect the signal mode outside the cavity (that is, we do not have
to take the signal field outside the cavity), but we couple it to a
meter field which is successively detected, thus overcoming the smearing effect
introduced by the direct detection of the signal~\cite{kn:dmt}. Moreover, there
is no need of a smoothing procedure, since we are interested in the marginal
probability distribution $W(x_s)$ which is directly related to $\tilde{W}(x_m)$
through Eq.~(\ref{eq:double}). In order to probe the full state of the signal
field, however, we would need to measure the probability distribution
$\tilde{W}(x_m)$ for various values of the
phase~\cite{kn:vog,kn:smith,kn:beck,kn:natu,kn:dmp,kn:gmd,kn:sch}.

\section{No measurement without a measurement}
\label{proof}

In this section we show that if a QND measurement performed on the
signal does not alter the probability density of the measured
observable, then the measurement process does not provide any information
about the measured observable itself. In order to keep the paper
self-contained, we prove this conclusion for the model considered here,
but this argument holds true also in general, independently of the chosen
model~\cite{kn:oalt}.
The argument is the following:

Let $\hat{\rho}_s(0)=|\psi_s(0)\rangle\langle\psi_s(0)|$ be the initial
density matrix of the signal, and $\hat{x}_s$ the measured observable,
with $\hat{x}_s|x_s\rangle=x_s|x_s\rangle$. The initial probability density
one would like to preserve is
$W_s^0(x_s)=\langle x_s|\hat{\rho}_s(0)|x_s\rangle$, and we are interested
in a QND measurement of $\hat{x}_s$. To this end, the signal is correlated
to a meter which is initially in a certain state $|\psi_m\rangle$, and
eventually a measurement is performed on the meter to yield the inferred
measurement result $\bar{x}_m$. The measurement is then completely
described~\cite{kn:sum} by the probability-amplitude operator
\begin{equation}
\hat{Y}(\hat{x}_s,\bar{x}_m)=
\langle\bar{x}_m|\hat{U}(\hat{x}_s,\hat{x}_m)|\psi_m\rangle\;,
\label{eq:probamp}
\end{equation}
which accounts
for the three stages of this measurement: preparation of the meter in the
state $|\psi_m\rangle$, interaction between the meter and the signal to
be measured through the unitary operator $\hat{U}(\hat{x}_s,\hat{x}_m)$
[see Eqs.~(\ref{eq:cohami}) and (\ref{eq:evol})], and projection
of the resulting entangled state onto the meter state $|\bar{x}_m\rangle$.
The QND condition~\cite{kn:sum} for a back-action evading measurement
then reads
\begin{equation}
[\hat{Y}(\hat{x}_s,\bar{x}_m),\hat{x}_s]=0\;,
\label{eq:qndcon}
\end{equation}
which means that $\hat{x}_s$ and $\hat{Y}$
share the same eigenstates:
\begin{mathletters}
\label{eq:eigen}
\begin{eqnarray}
\hat{Y}(\hat{x}_s,\bar{x}_m)|x_s\rangle&=&Y(x_s,\bar{x}_m)|x_s\rangle\;,
\label{eq:eigeny} \\
\hat{Y}^{\dagger}(\hat{x}_s,\bar{x}_m)|x_s\rangle&=&Y^{\ast}(x_s,\bar{x}_m)
|x_s\rangle\;.
\label{eq:eigenyd}
\end{eqnarray}
\end{mathletters}

After a measurement which gives the result $\bar{x}_m$, the system is
therefore described by the density matrix
\begin{equation}
\hat{\rho}_s=\frac{1}{W(\bar{x}_m)}\hat{Y}(\hat{x}_s,\bar{x}_m)
\hat{\rho}_s(0)\hat{Y}^{\dagger}(\hat{x}_s,\bar{x}_m)\;,
\label{eq:densmatr}
\end{equation}
where
\begin{eqnarray}
W(\bar{x}_m)&=&
{\rm Tr}_s[\hat{Y}(\hat{x}_s,\bar{x}_m)\hat{\rho}_s(0)\hat{Y}^{\dagger}
(\hat{x}_s,\bar{x}_m)]
\nonumber \\
&=&\int dx_s\,\langle x_s|\hat{Y}(\hat{x}_s,\bar{x}_m)\hat{\rho}_s(0)
\hat{Y}^{\dagger}(\hat{x}_s,\bar{x}_m)|x_s\rangle
\label{eq:pro}
\end{eqnarray}
is the probability to obtain the result $\bar{x}_m$. Now, the probability
density of the measured observable after the measurement is given by
\begin{eqnarray}
W_s(&x_s&)=\langle x_s|\hat{\rho}_s|x_s\rangle
\nonumber \\
&=&\frac{1}{W(\bar{x}_m)}\langle x_s|\hat{Y}(\hat{x}_s,\bar{x}_m)\hat{\rho}_s(0)
\hat{Y}^{\dagger}(\hat{x}_s,\bar{x}_m)|x_s\rangle\;.
\label{eq:proden}
\end{eqnarray}
Applying the QND condition~(\ref{eq:qndcon}) and~(\ref{eq:eigen}) we obtain
\begin{equation}
W_s(x_s)=\frac{1}{W(\bar{x}_m)}|Y(x_s,\bar{x}_m)|^2 W_s^{(0)}(x_s)\;.
\label{eq:finpro}
\end{equation}

If we require that this probability density does not change due to the
measurement process, $W_s(x_s)=W_s^{(0)}(x)$, then it must be that
\begin{equation}
|Y(x_s,\bar{x}_m)|^2=W(\bar{x}_m)\;.
\label{eq:ugua}
\end{equation}
However, $W(\bar{x}_m)$ is not a function of $x_s$ (the eigenvalues of the
measured observable) and therefore also the eigenvalues $Y(x_s,\bar{x}_m)$
of $\hat{Y}(\hat{x}_s,\bar{x}_m)$ are independent of $x_s$. Since the operator
$\hat{Y}$ describes the measurement process, if its eigenvalues are
independent of the eigenvalues of $\hat{x}_s$, the measurement obviously
gives no information about $\hat{x}_s$, unless the measured state is an
eigenstate of the measured observable.

\section{Conclusions}
\label{conclu}

In this paper we have proposed a method to measure the quadrature probability
distribution (or, more generally, the full quantum state) of a single mode
of the electromagnetic field inside a cavity. It is based on indirect
homodyne measurements performed on a meter field which is coupled to the
signal field via a QND interaction Hamiltonian. We have named this procedure
``endoscopic tomography'' because (i) it does not require (in contrast to
Refs.~\cite{kn:smith,kn:beck,kn:natu}) to take the field out of the cavity,
just as in
``quantum state endoscopy''~\cite{kn:bard}, where a beam of two-level atoms
is used as a probe; (ii) tomographic measurements performed (by balanced
homodyne detection) on the meter mode allow us to reconstruct the marginal
probability distribution of the signal variable or even the full quantum state.

We have computed the entangled (signal-meter) state which arises during the
evolution under the QND Hamiltonian, and evaluated the conditional signal state
(given that a measurement on the meter has provided a certain result). Then,
we have concentrated ourselves on two special cases, namely, in phase and out
of phase measurements. We have shown that in the first case the shape of the
Wigner function of the signal is not changed by the measurement, but also
that such a measurement does not provide any information on the signal state.
In the second case, however, we can get information about the signal, but
its initial state is changed due to the measurement performed on the meter:
in this case, preparing the signal field again in the same state after each
measurement, balanced homodyne detection of the meter mode allows the
reconstruction of the original signal state. Finally, we have given an argument
according to which the results we have found in our model are rather general:
a QND measurement which leaves unchanged the probability distribution of the
system observable does not provide any information on the signal state.

\acknowledgments

We gratefully thank O.~Alter, F.~Harrison, J.~H.~Kimble and K.~M\"olmer
for fruitful discussions. 
This work was partially supported by the European Union (TMR programme),
the Deutsche Forschungsgemeinschaft, the Land Baden-W\"urttemberg,
and INFM through the 1997 PRA ``Cat''.

\appendix
\section{Displacement of the meter state}
\label{dispmeter}

In this appendix we calculate the state 
\begin{equation}
|\bar{\psi}_m \rangle \equiv \exp [-i 2 \sigma tx_s \hat{x}_m (\phi)] |\psi_m
\rangle \equiv \hat{D} (\beta)|\psi_m \rangle 
\label{eq:opstate}
\end{equation}
which results from the application of the operator
$\hat{D} (\beta) \equiv \exp
[-i \beta \hat{x}_m (\phi)]$ on the meter state $|\psi_m \rangle$
with $\beta = 2 \sigma tx_s$. When we use the representation
\begin{equation}
|\psi_m \rangle = \int\limits_{-\infty}^\infty dx_m \psi_m (x_m; \theta)
| x_m(\theta) \rangle
\label{eq:repr}
\end{equation}
in quadrature states $|x_m (\theta) \rangle$ at the angle $\theta$, the state 
$|\bar{\psi}_m \rangle$ reads
\begin{equation}
|\bar{\psi}_m \rangle = \int \limits_{-\infty}^\infty dx_m
\psi_m (x_m ; \theta) \exp [-i\beta \hat{x}_m (\phi)]|x_m(\theta)\rangle\;.
\label{eq:stread}
\end{equation}
We recall the result 
\begin{eqnarray}
&\exp& [-i\beta \hat{x} (\phi)]| x_m (\theta)\rangle=
\nonumber \\
&\exp& \left\{ -i\left[\frac{3\beta^2}{4} \sin \left[2(\theta - \phi)\right]
+ \beta x_m \cos (\theta-\phi)\right]\right\}
\nonumber \\
&\times&|[x_m + \beta \sin (\theta - \phi)] (\theta)\rangle     
\label{eq:reres}
\end{eqnarray}
derived in Appendix~\ref{quad} and find 
\begin{eqnarray}
| \bar{\psi}_m \rangle &=& \int\limits_{-\infty}^\infty dx_m \psi_m
(x_m; \theta)
\nonumber \\
\times &\exp& \left\{-i \left[\frac{3\beta^2}{4} \sin\left[2(\theta-\phi)\right]
+ \beta x_m \cos(\theta-\phi)\right]\right\}
\nonumber \\
&\times& |[x_m +\beta \sin (\theta - \phi)] (\theta) \rangle\;,
\end{eqnarray}
which after introducing the integration variable
$\bar{x}_m \equiv x_m + \beta\sin(\theta-\phi)$ reads 
\begin{eqnarray}
|\bar{\psi}_m \rangle&=&\int\limits_{-\infty}^\infty
d\bar{x}_m \psi_m\left[\bar{x}_m-\beta\sin(\theta-\phi);\theta\right]
\nonumber \\
\times &\exp& \left\{-i\left[\frac{\beta^2}{4}\sin\left[2(\theta - \phi)\right]
+\beta\bar{x}_m\cos(\theta - \phi)\right]\right\}
\nonumber \\
&\times& |\bar{x}_m(\theta)\rangle\;.
\label{eq:finres}
\end{eqnarray}
Hence the meter wave function gets displaced and experiences a phase shift.

\section{Displacement of a quadrature state}
\label{quad}

In this appendix we derive the relation 
\begin{eqnarray}
e^{-i\beta \hat{x} (\theta)} |x(\theta')\rangle &=& 
\exp\left[i\varphi(x; \beta, \theta' - \theta)\right]
\nonumber \\
&\times& |\left[x+\beta\sin(\theta'-\theta)\right](\theta') \rangle  
\label{eq:relation}
\end{eqnarray}
for the c-number $\beta$ and the quadrature operator 
\begin{equation}
\hat{x} (\theta) \equiv \frac{1}{\sqrt{2}} (\hat{a} e^{-i\theta}
+ \hat{a}^{\dagger} e^{i\theta})\;.
\label{eq:quad}
\end{equation}
Here $\hat{a}$ and $\hat{a}^{\dagger}$ denote the annihilation and creation 
operators, respectively, with 
\begin{equation}
[\hat{a}, \hat{a}^\dagger] = 1\;.
\label{eq:comm}
\end{equation}
Note that according to Eq.~(\ref{eq:relation}) the action of the exponential
of the quadrature operator $\hat{x} (\theta)$ at the angle $\theta$ on a
quadrature eigenstate $|x (\theta')\rangle$ at angle $\theta'$ yields, apart
from the phase
\begin{equation}
\varphi(x; \beta, \theta' - \theta)=-\frac{3\beta^2}{4}\sin[2(\theta'- 
\theta)] - \beta x \cos (\theta' - \theta)\;,
\label{eq:phase}
\end{equation}
again a quadrature eigenstate at the angle $\theta'$, but with the
eigenvalue
\begin{equation}
x' \equiv x + \beta \sin (\theta' - \theta)\;.
\end{equation}
To prove Eq.~(\ref{eq:relation}) we first express the operator
$\hat{x} (\theta')$, Eq.~(\ref{eq:quad}), in quadrature operators
\begin{equation}
\hat{x} (\theta') = \frac{1}{\sqrt{2}} (\hat{a} e^{-i \theta'} +
\hat{a}^{\dagger} e^{i \theta'})
\label{eq:opquad}
\end{equation}
and
\begin{eqnarray}
\hat{p}(\theta') &\equiv& \hat{x} (\theta' + \pi/2)
\nonumber \\ 
&=& \frac{1}{\sqrt{2}i} (\hat{a} e^{-i \theta'} - \hat{a}^\dagger 
e^{i \theta'})
\label{eq:pquad}
\end{eqnarray}
at the angle $\theta'$. After minor algebra we find using these expressions
the relation 
\begin{equation}
\hat{x}(\theta)=\cos(\theta'-\theta)\hat{x}(\theta')-\sin(\theta'
-\theta)\hat{p}(\theta')\;.
\end{equation}
The Baker-Hausdorff relation~\cite{kn:lou}
\begin{equation}
e^{\hat{A}+\hat{B}}=e^{\hat{A}}e^{\hat{B}}e^{-\frac{1}{2}[\hat{A},\hat{B}]}
\label{eq:baker}
\end{equation}
for two operators $\hat{A}$ and $\hat{B}$ with
$[\hat{A},[{\hat{A},\hat{B}}]]=[\hat{B},[{\hat{A} ,\hat{B}}]]=0$ yields
\begin{eqnarray}
&\exp& [-i\beta \hat{x} (\theta)] = \exp [-i\beta \cos (\theta' - \theta) 
\hat{x} (\theta')]
\nonumber \\
&\times&\exp[i\beta \sin (\theta' - \theta) \hat{p} (\theta')]
\nonumber \\ 
&\times&\exp \left\{-i \frac{\beta^2}{4} \sin [2(\theta' - \theta)]\right\}\;,
\label{eq:expexp}
\end{eqnarray}
where we have made use of $[\hat{x}(\theta'),\hat{p}(\theta')]=i$,
following from Eqs.~(\ref{eq:comm}), (\ref{eq:opquad}), and (\ref{eq:pquad}).

Recalling the displacement property 
\begin{equation}
e^{iy \hat{p}} | x\rangle = | x+y \rangle
\label{eq:disp}
\end{equation}
of the momentum operator $\hat{p}$, we find using the representation
Eq.~(\ref{eq:expexp}) of the operator $\hat{x}(\theta)$ the expression 
\begin{eqnarray}
& &\exp [-i\beta \hat{x} (\theta)]| x (\theta') \rangle = \exp
\left\{-i \frac{\beta^2}{4}\sin[2(\theta'-\theta)]\right\}
\nonumber \\
& & \qquad \times \exp[-i \beta \cos (\theta' -\theta) \hat{x} (\theta')]
\nonumber\\
& & \qquad \times |[x + \beta \sin (\theta' - \theta)] (\theta') \rangle\;,
\label{eq:expr}
\end{eqnarray}
or
\begin{eqnarray}
&\exp& [-i \beta \hat{x} (\theta)]| x (\theta') \rangle = \exp
\left\{-i \frac{3\beta^2}{4} \sin [2 (\theta' - \theta)]\right\}
\nonumber \\
&\exp&[-i\beta x\cos (\theta' - \theta)]
|[x + \beta \sin (\theta' - \theta)] (\theta') \rangle
\label{eq:res}
\end{eqnarray}
which is the result Eq.~(\ref{eq:relation}).

\end{document}